\documentclass[11pt, a4paper, oneside]{article}
\usepackage{cite}
\usepackage{graphicx}
\usepackage{epsfig}
\usepackage{epstopdf}
\usepackage{subfigure}
\usepackage{amssymb,amsmath}
\usepackage[a4paper=true,pagebackref=false]{hyperref}
\usepackage[scale=0.8]{geometry}
\hypersetup{colorlinks = true, linkcolor = cyan, anchorcolor = green, citecolor = red, filecolor = red, pagecolor = magenta, urlcolor = blue}

\title{Qualitative study of anisotropic Rastall cosmologies}

\author{A Singh$^{1}$\footnote{Corresponding author: A Singh, Email: ashuverse@gmail.com}, A Pradhan$^{2}$ \\ \\
$^{1}$Department of Applied Mathematics,\\ Jabalpur Engineering College, Jabalpur, Pin - 482011, India\\ \\ $^{2}$Centre for Cosmology,
Astrophysics and Space Science,\\ GLA University, Mathura, Pin - 281406, India}

\date{}
\begin{document}
\maketitle

\begin{abstract}
We perform dynamical analysis of anisotropic Rastall cosmologies and, in particular Kantowski-Sachs, locally rotationally symmetric (LRS)
Bianchi I and LRS Bianchi III cosmologies. Using dynamical system techniques, a qualitative analysis of cosmological equations yield physically interesting cases which are in agreement with observations such as de Sitter, accelerating, stable attractors are isotropic. 
Features such as isotropization and cosmological bounce are discussed in detail.
\end{abstract}

%\noindent\hrulefill
%\newpage
\section{Introduction}
\label{intro}
Rastall's theory is a non-Lagrangian theory of gravity, where the divergence-free property of Einstein's tensor leading to the usual 
conservation law for energy-momentum source of Einstein's gravity is replaced by a non-zero divergence of the energy-momentum tensor of matter 
field \cite{pr1972}. In this theory, departure from usual General relativity conservation law is parameterized by a dimensionless parameter 
$\lambda$. Rastall remarked that the conservation law ${T^{ij}}_{;i}=0$ may not hold true in curved space-time, hence the modification 
of conservation law may be justified \cite{pr1972}. The particle creation phenomena in cosmology also leads to the violation of classical 
conservation law, and, in this sense, the modification of conservation law in Rastall theory may be viewed as a kind of classical formulation 
of that quantum phenomena, since the violation of the conservation of energy-momentum tensor is connected with the curvature 
\cite{gw1977,ndb1982,lh1987,cem2012}. Therefore, the modification of General relativity by considering a non-minimal coupling between geometry 
and matter field leads to the non-zero divergence of energy-momentum tensor in this theory. The violation of the energy-momentum conservation equation 
is not specific to the Rastall gravity and this violation has also provided motivations for constructing a possible Lagrangian 
formulation \cite{5a,5b}. It is worthwhile to mention that Rastall gravity is different from General relativity \cite{21}. In literature, 
Darabi et al. \cite{14} have illustrated that Rastall gravity is a form of modified gravity different from General relativity; on the other 
hand, Visser has claimed that the Rastall gravity is equivalent to General Relativity \cite{r21a}.  
Various aspects of this theory including theoretical and observational ones have been reported in literature 
\cite{r21a,mc2010,6,9,12,13,14,sm2020,23b,21,51a,23z,23y,23x,01,1a}.\\
Most of the governing equations in cosmological models are non-linear, and thus dynamical system analysis may be used to describe the qualitative 
behavior of the model. In the dynamical system analysis of cosmological model, the fixed points of the autonomous system governing different epochs of 
the universe are identified for qualitative analysis of solutions. In this paper, we study the system of autonomous differential equations generated 
from the governing equations of the Rastall theory in anisotropic space-time. Spatially homogeneous cosmological models with anisotropies 
in relativistic cosmology yield richer dynamical structure. Despite its complexity, the cosmological model remains simple enough to yield 
numerical and analytical results. These homogeneous and anisotropic models allow us to investigate the issues like models' behavior as they 
approach space-time singularities, why the present-day universe appears homogeneous and isotropic etc. Bianchi geometries generalize the 
isotopic feature of Friedmann-Robertson-Walker (FRW) geometries.  Bianchi spacetimes are characterized by different degrees of anisotropies, 
but retain the homogeneity properties. Homogeneous and anisotropic universe may have significant importance in early universe modeling 
\cite{montani,crf,lpnrgv}. During early times, universe geometry might be different from completely homogeneous and isotropic space-times
since spatially homogeneous models may be unstable \cite{61,62}. Under the light of Planck observations \cite{planck2013} also, the anisotropic 
geometries have gained a lot of interest due to the small anomalies appearing in the observations. Rastall gravity theory explain different 
aspects of universe evolution \cite{cem2012,21,6,12,23b,23z,23y}. Therefore, it would be interesting to explore the full phase of possibilities 
with the anisotropic, homogeneous Bianchi and Kantowski-Sachs space-times in Rastall gravity. The dynamical system method allows us to 
identify the general behavior of cosmological solutions associated with the present model and can be used to determine the presence and 
stability of solutions of cosmological interest, such as those with de Sitter phases or with radiation, as well as matter-dominated phases. 
This method has been applied to a wide range of cosmological models originating from different gravity theories, see for example 
\cite{9,ejcop,mggfr,Carloni,dspks2006,jdb2013,crf,lpnrgv,71,64,73,smsc,Mg2016,lng,21,mas2020,4,1a,10,01,61,62,63} and references therein. 
Khyllep and Dutta have investigated that the dynamics of the Rastall model in flat FRW space-time resembles the $\Lambda$ cold 
dark matter model at the background level during late times \cite{21}. The Rastall gravity with flat FRW spacetime may yield simple 
power-law and, bouncing solutions with ekpyrotic evolution in the contracting era \cite{sm2020}. By using the quadratic equation of state signifying 
high energy state, the stability of bouncing solutions in the Rastall theory framework has been investigated by 
Silva et al. \cite{9}. Using dynamical system method in Rastall gravity, the oscillating solutions signifying cyclic universe evolution 
may be realized in homogeneous and isotropic space-time having flat, as well as closed spatial sections \cite{01,1a}. The two-dimensional 
state-space in flat FRW spacetime with Rastall gravity may always be Darboux integrable, and global dynamics of the differential equations 
may be classified into non-topological equivalent phase portraits \cite{23x}.\\
It is convenient to write the governing equations of expansion scalar, shear, and $3$-curvature in anisotropic cosmologies as an autonomous system 
of first order non-linear differential equations. These equations may be used to perform a local analysis to characterize the stability of fixed 
points corresponding to the specific cosmological solutions. The stability of each fixed point is determined by the eigenvalues of the Jacobian matrix 
evaluated at the corresponding fixed point. The points are stable (unstable) if all the eigenvalues have negative (positive) real parts, or saddle 
for eigenvalues with real parts of different signs. Our aim in this paper is to study the qualitative behavior of LRS Bianchi-I (LRS BI), 
LRS Bianchi-III (LRS BIII) and Kantowski-Sachs (KS) space-times in the Rastall gravity framework. We use this analysis to study the cosmological 
features of the resulting system.\\
The paper is organized as follows: in section \ref{section2}, we write the cosmological equations. In section \ref{KS} and \ref{LRSB}, we 
perform the dynamical system analysis of Kantowski-Sachs and LRS BI with LRS BIII model, respectively. In section \ref{GI}, we discuss issues 
like late-time isotropization and bouncing behavior of the resulting cosmologies. In section \ref{conclusion}, we summarize our results.
\section{Cosmological equations}
\label{section2}

For the anisotropic metric and choice of co-moving fluid vector $u^i$, the modified Einstein's field equations in terms of 
propagation equations for the expansion $\Theta$, shear $\sigma$, and three-curvature scalar ${}^3R$ may be given in the Rastall gravity. The field equations are given as 
\cite{pr1972,mc2010}
\begin{eqnarray}
R_{ij}-\frac{1}{2}(1-2k\lambda)Rg_{ij}=kT_{ij} 
\label{eq1}
\end{eqnarray}
where $R_{ij}, R, g_{ij}, T_{ij}$ are Ricci tensor, Ricci scalar, metric tensor, stress-energy tensor of fluid respectively. \\
In a generic fluid flow, geometrical quantities such as expansion scalar, shear, and vorticity may be non-zero, so the combination of 
effects like volume change, distortion and rotation may occur in a relativistic model. The vector field $u$ is said to be irrotational 
if the vorticity is zero. And if the congruence $u$ is irrotational, formulae for the curvature of $3-$spaces orthogonal to the congruence 
may be obtained. So, in the Rastall gravity, one may write the Gauss-Codazzi, shear propagation, three-curvature propagation, and the Raychaudhuri 
like equation as \cite{sm2020}:
\begin{eqnarray}
\frac{2}{3}\Theta^2+{}^3{R}-2\sigma^2=2k\left(\rho-\frac{k\lambda}{4k\lambda-1}(\rho-3p) \right) \nonumber\\
\dot{\sigma}+\Theta\sigma-\frac{1}{2\sqrt{3}}({}^3{R})=0 \nonumber \\
{}^3{\dot{R}}+{}^3{R}\frac{2}{3}\Theta-{}^3{R}\frac{2}{\sqrt{3}}\sigma=0 \nonumber\\
\dot{\Theta}+2\sigma^2+\frac{1}{3}\Theta^2=k\left(\rho-\frac{1}{2}\frac{2k\lambda-1}{4k\lambda-1}(\rho-3p) \right)
\label{eq2}
\end{eqnarray} 
The homogeneous and anisotropic locally rotationally symmetric metric may be given by \cite{62}
\begin{equation}
ds^2=dt^2-{a_{1}}^2dr^2-{a_{2}}^2(d\theta^2+ f^2 (\theta) d\phi^2) 
\label{eq3}
\end{equation}
where $a_{1}(t),a_2(t)$ are directional scale factors. The above metric (\ref{eq3}) may reduce to the KS, LRS BIII and LRS BI metric for 
$f(\theta)= \sin \theta$, $f(\theta)= \sinh \theta$ and $f(\theta)=\theta$, respectively.  For metric (\ref{eq3}), we may also write 
$\Theta=3H=\frac{\dot a_{1}}{a_{1}}+\frac{2\dot a_{2}}{a_{2}}$, $\sigma=\frac{1}{\sqrt{3}}\left( \frac{\dot a_{1}}{a_{1}}-\frac{\dot a_{2}}{a_{2}}\right) $, 
${}^3{R}=\frac{2\kappa}{{a_2}^2}$, where $H$ is mean Hubble parameter \cite{sm2020}. The deceleration parameter may be defined as 
$q=-1-\frac{3\dot{\Theta}}{\Theta^2}$; indicates the rate at which the expansion of universe is slowing down. Universe is accelerating (decelerating) 
for $q<0$ ($q>0$). Eternal acceleration scenario is achieved if $q<0$ during the complete evolution history of universe. The de Sitter, 
accelerating power-law expansion can be achieved for $q =-1$ and $-1 < q < 0$ respectively. For $q<-1$, super-exponential expansion scenario 
exists \cite{9a,gps2018,A2020,1,2}. It have been confirmed by observational results that the universe is currently expanding with acceleration 
and it has been decelerating in the past \cite{Pl2018}. We consider a perfect fluid with barotropic equation of state $p=f(\rho)$ in our present 
setting and keep our discussion general for the moment. 
\section{Kantowski-Sachs cosmology}
\label{KS}
\begin{figure}[h!]
  \centering
  \includegraphics[width=.375\textwidth]{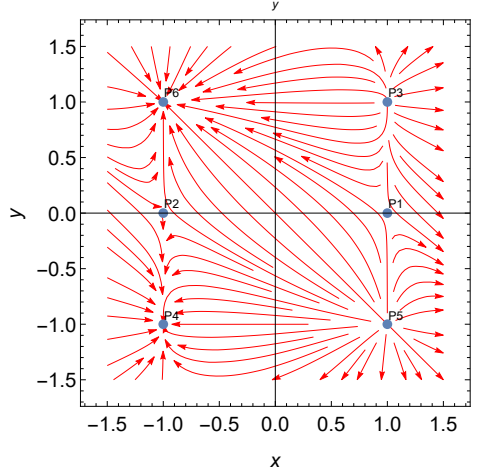}\hfill
  \caption{Phase space $(x,y)$ in the sub-manifold $u=0$}
  \label{figure1}
  \end{figure}

\begin{figure}[h!]
 	\centering
 	\subfigure[]{\includegraphics[width=0.375\textwidth]{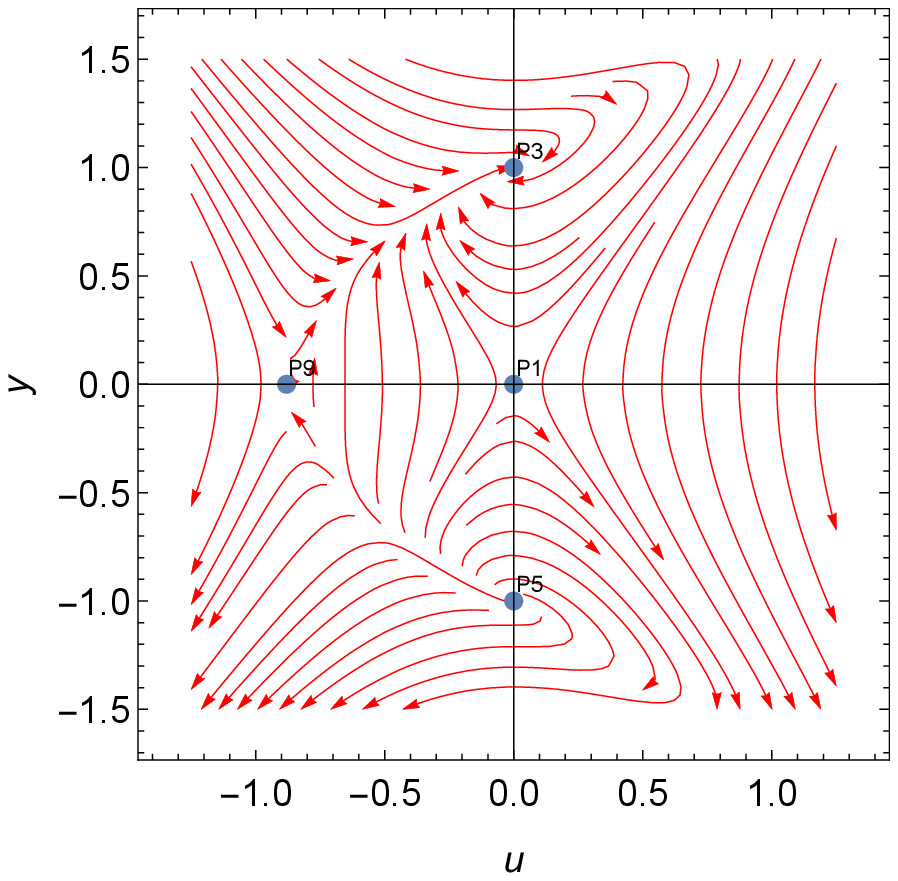}} \hfill
 	\subfigure[]{\includegraphics[width=0.375\textwidth]{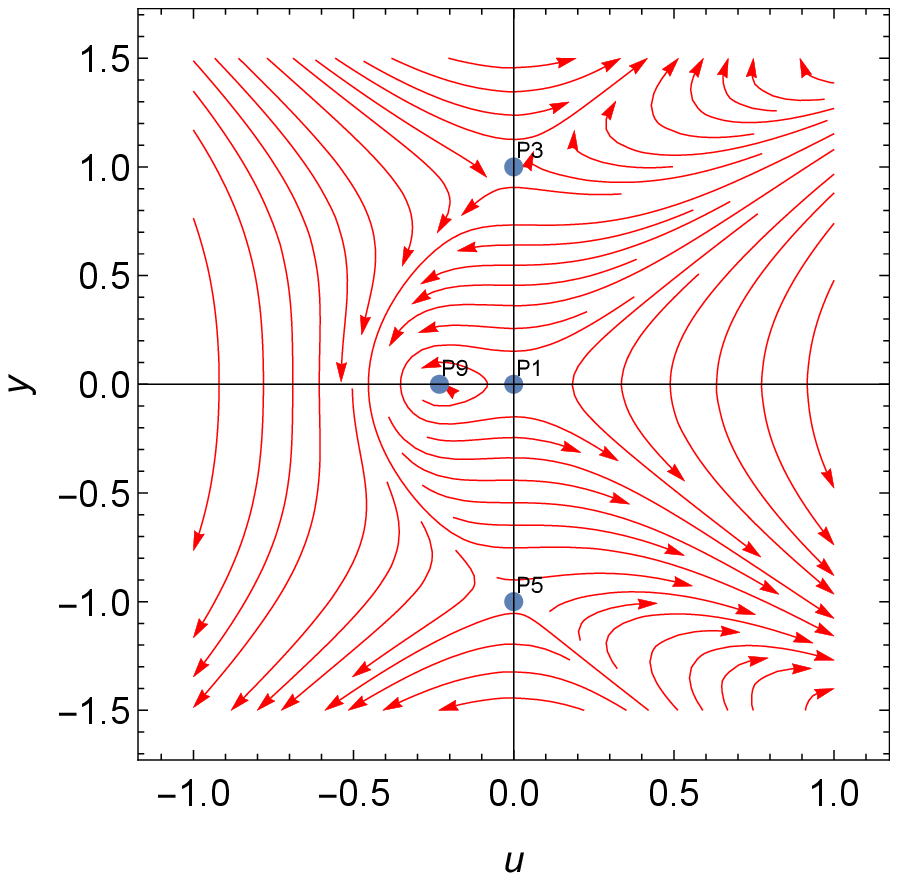}} 
 	\subfigure[]{\includegraphics[width=0.375\textwidth]{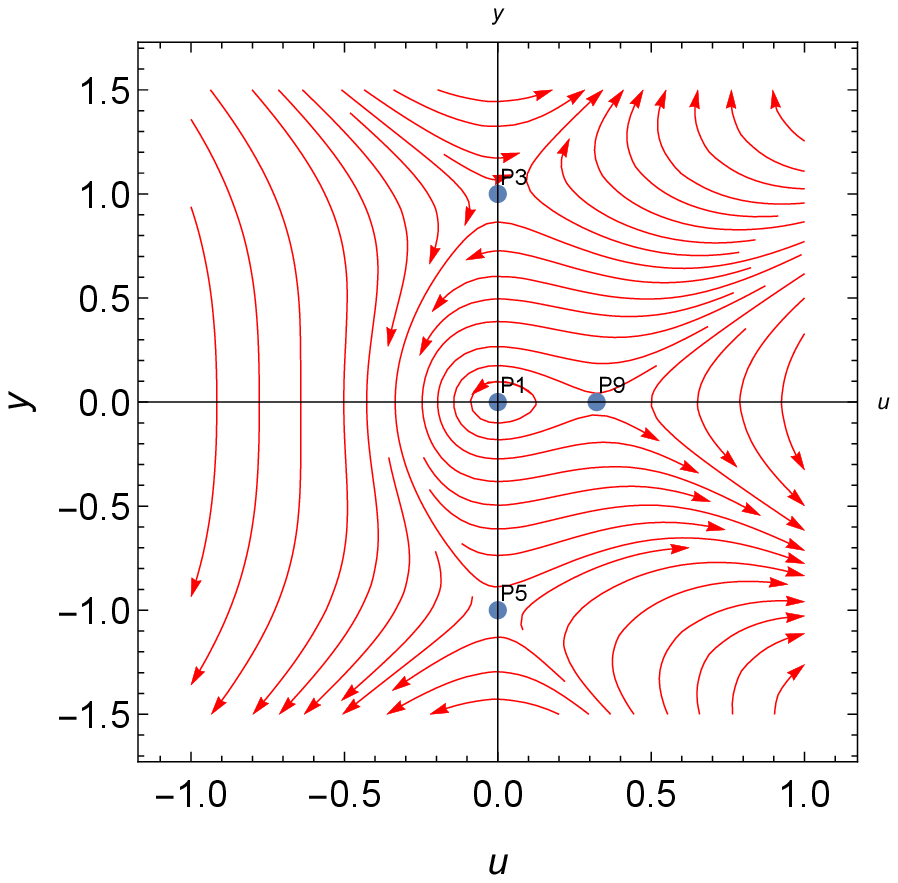}}\hfill
 	\subfigure[]{\includegraphics[width=0.375\textwidth]{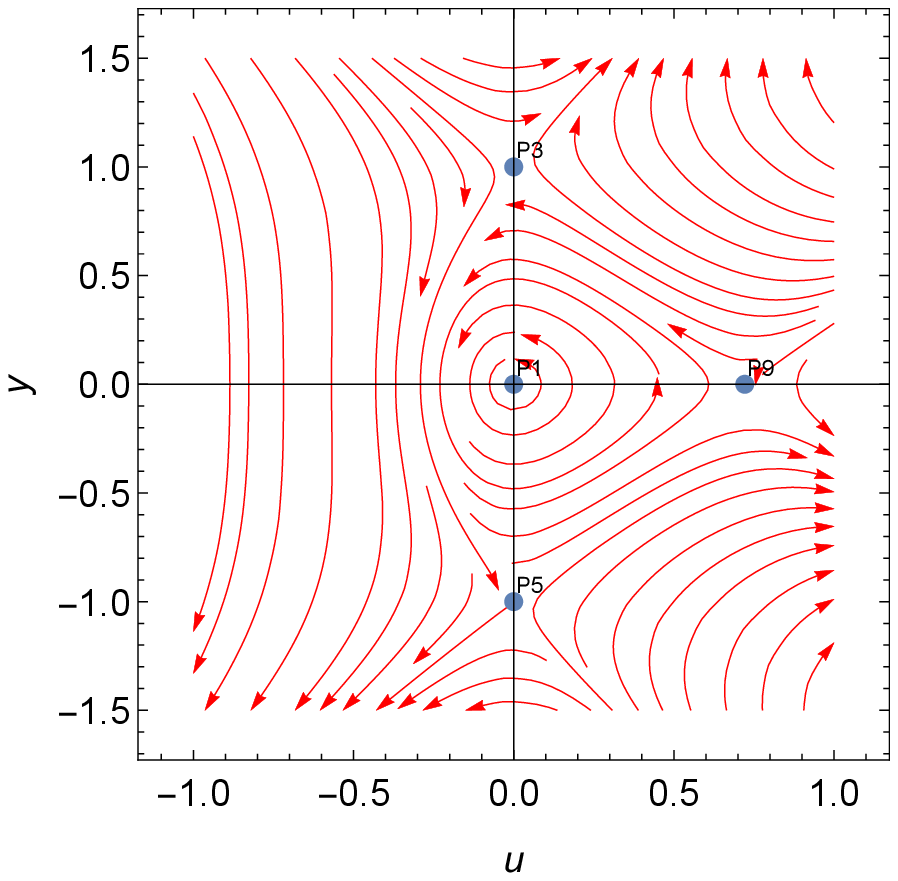}}
 	
 	\caption{Phase space $(u,y)$ in the sub-manifold $x=1$ for (a) $A<-2$, (b) $-2<A<-\frac{4}{3}$, (c) $-\frac{4}{3}<A<-\frac{2}{3}$ 
 	and (d) $A>-\frac{2}{3}$ respectively \label{figure2}}
 \end{figure}

\begin{figure}[h!]
	\centering
	\subfigure[]{\includegraphics[width=0.375\textwidth]{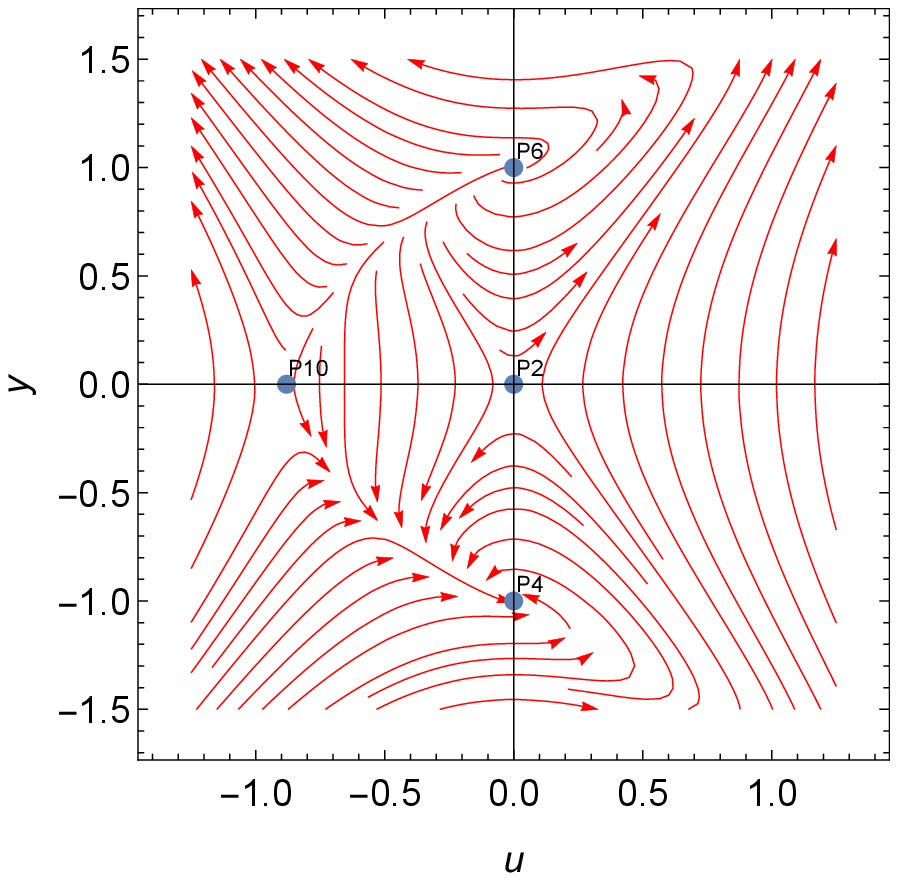}} \hfill
	\subfigure[]{\includegraphics[width=0.375\textwidth]{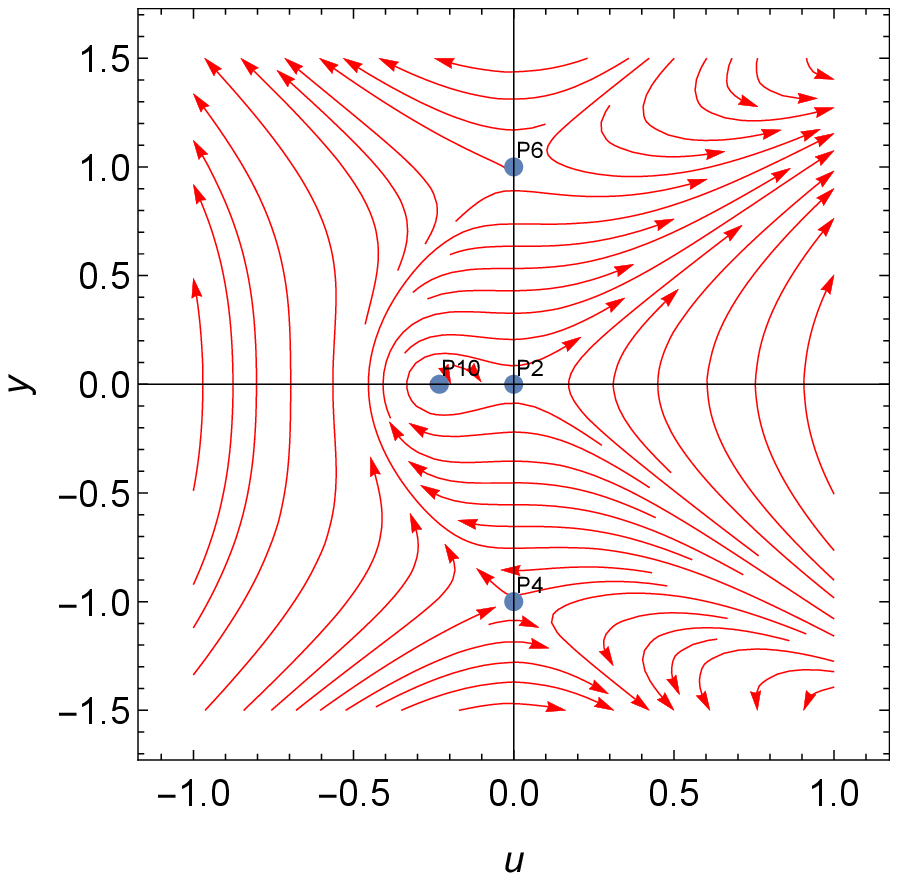}} 
	\subfigure[]{\includegraphics[width=0.3775\textwidth]{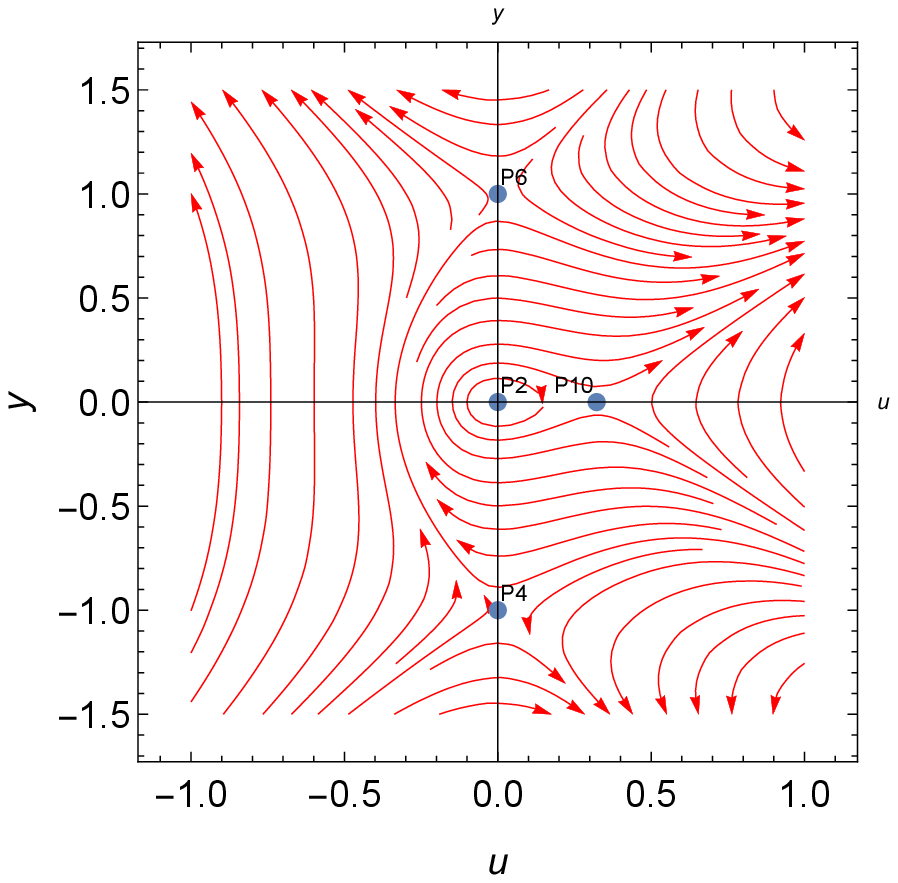}}\hfill
	\subfigure[]{\includegraphics[width=0.375\textwidth]{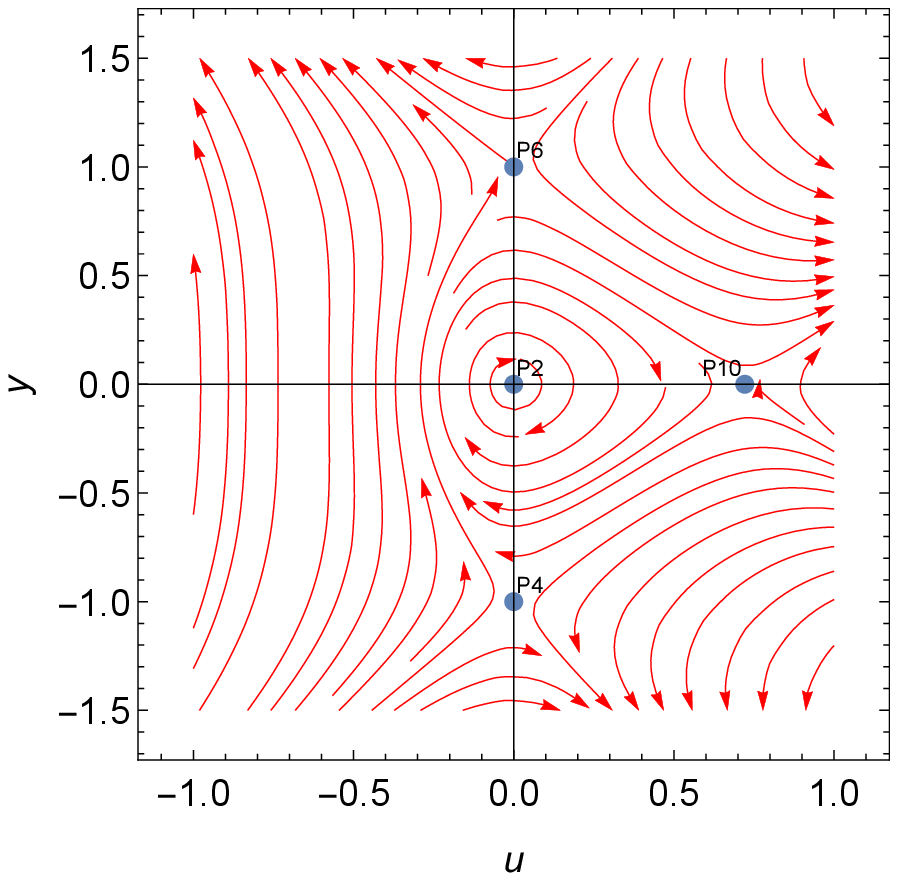}}
		\caption{Phase space $(u,y)$ in the sub-manifold $x=-1$ for (a) $A<-2$, (b) $-2<A<-\frac{4}{3}$, 
		(c) $-\frac{4}{3}<A<-\frac{2}{3}$ and (d) $A>-\frac{2}{3}$ respectively \label{figure3}}
\end{figure}

In the Friedmann equation, we define the non-negative quantity $D^2=\frac{\Theta^2}{9}+\frac{^3R}{6}$. Since $D$ is real-valued and strictly 
positive, it provides a monotonically increasing time variable. We define the dynamical variables 
\begin{equation}
x^2=\frac{\Theta^2}{9D^2}, \ \ y^2=\frac{\sigma^2}{3D^2}, \ \ z=\frac{{}^3R}{6D^2}, \ \  u=\frac{k(\rho+p)}{3D^2}, \ v=\frac{k\rho}{3D^2}
\label{eq4}
\end{equation} 
We take $\tau=Dt$ and `prime' denotes derivative with respect to $\tau$. Also, $-1\leq x\leq 1$, $-1\leq y\leq 1$, $0\leq z\leq 1$, $-1\leq u\leq 1$ 
and $v\geq 0$. From definition of $D$ and Gauss-Codazzi equation (\ref{eq2}), we have $x^2+z=1$ and $v=3k\lambda u+(1-4k\lambda)(1-y^2)$. In terms of 
dynamical variables (\ref{eq4}), the deceleration parameter $(q)$, matter density parameter ($\Omega_m$) may be written as  
\begin{eqnarray}
q=\frac{1}{x^2}\left(1+y^2-\frac{3}{2}u \right), \quad 
\Omega_m=\frac{\rho}{3H^2}=\frac{3k\lambda u+(1-4k\lambda)(1-y^2)}{kx^2}
\label{eq5}
\end{eqnarray}
From the sign of $x$, we may determine whether the universe is expanding or contracting. $y$ determines the anisotropy level with $y=0$ corresponding 
to the isotropization. $z$ is related to the curvature density parameter, and $z=0$ corresponds to a flat universe. Due to auxiliary relations between $x,z$ 
and $u,v,y$, the autonomous system is $3$-dimensional and may be written as
\begin{eqnarray}
x'=\frac{1}{2}(x^2-1)\left(2(1+xy+y^2)-3u \right) \nonumber \\
y'=(1-x^2)(1-y^2)+xy\left(y^2-1-\frac{3}{2}u \right)\nonumber \\
u'=u\left(2y(x^2-1)+x(4+3A-3u+2y^2)\right)
\label{eq6}
\end{eqnarray}
along with the evolution equations of auxiliary variables given by
\begin{eqnarray}
D'=D\left(-x^3(1+q)-xz+yz \right) \nonumber \\
z'=2z\left(x^3(1+q)+(x-y)(z-1) \right)\nonumber \\
v'=\frac{3Aux}{1+f'(\rho)}+2v\left(x^3(1+q)+xz-yz \right)
\label{eq7}
\end{eqnarray}
where we use $p=f(\rho)$ and $f'(\rho)=\frac{df}{d\rho}$. The areas of phase space that evolves themselves under the dynamics may be termed 
an invariant set. $x=\pm 1$ will  correspond to $z=0$, with $x>0$ and $x<0$ corresponds to the expanding and contracting universe, respectively. 
For $p=\alpha\rho$, we find the fixed points of system by setting $x'=0,y'=0,u'=0$. For the above system (\ref{eq6}), there are $12$ fixed points. 
Various details about these fixed points have been listed in Table (\ref{table1}). From $\frac{3\dot{\Theta}}{\Theta^2}=-(1+q)$, we may have 
$a(t)\propto t^{\frac{1}{1+q_0}}$, where $a(t)$ is mean scale factor related to volume scale factor $V$ by $V=a^3=a_1{a_2}^2$ and $q_0$ is 
the deceleration parameter value calculated at the fixed point. We write $A\equiv \frac{(1+\alpha)(1-4k\lambda)}{3k\lambda(1+\alpha)-1}$ and 
this notation is uniformly used in this manuscript. In terms of $A$, we may write $k\lambda=\frac{1+\alpha+A}{(4+3A)(1+\alpha)}$. It simply 
means that at $A=-\frac{4}{3}$ and $\alpha=-1$, the quantity $k\lambda$ may diverge. For $p=\alpha\rho$, the quantity $\frac{df}{d\rho}=\alpha$ 
may be termed as a measure for the classical stability of model. By using the classical stability criterion \cite{01} for $p=\alpha\rho$, we may 
constrain $\alpha$ as $0\leq \alpha\leq 1$. With $\alpha=-1$, the model will not be classically stable. We define the effective equation of 
parameter $(\gamma)$ in the present model as $\gamma=-1-\frac{2}{3}\frac{\dot{H}}{H^2}$. \\
 \begin{table}[h!]
 \begin{center}
 {\begin{tabular}{ccccccc}
 \hline\noalign{\smallskip}
 Point & $x$ & $y$ & $u$ & Existence & $q$ & $\Omega_m$ \\
 \noalign{\smallskip}\hline\noalign{\smallskip}
$P_1$ & $1$ & $0$ & $0$ & always & $1$ & $\frac{1-4k\lambda}{k}$ \\
$P_2$ & $-1$ & $0$ & $0$ & always & $1$ & $\frac{1-4k\lambda}{k}$ \\
$P_3$ & $1$ & $1$ & $0$ & always & $2$ & $0$\\
$P_4$ & $-1$ & $-1$ & $0$ & always & $2$ & $0$ \\
$P_5$ & $1$ & $-1$ & $0$ & always & $2$ & $0$ \\
$P_6$ & $-1$ & $1$ & $0$ & always & $2$ & $0$ \\
$P_7$ & $2$ & $-1$ & $0$ & always & $\frac{1}{2}$ & $0$ \\
$P_8$ & $-2$ & $1$ & $0$ & always & $\frac{1}{2}$ & $0$ \\
$P_9$ & $1$ & $0$ & $A+\frac{4}{3}$ & always & $-1-\frac{3A}{2}$ & $\frac{1+3Ak\lambda}{k}$ \\
$P_{10}$ & $-1$ & $0$ & $A+\frac{4}{3}$ & always & $-1-\frac{3A}{2}$ & $\frac{1+3Ak\lambda}{k}$ \\
$P_{11}$ & $\frac{-2}{4+3A}$ & $\frac{-2-3A}{4+3A}$ & $\frac{4(1+A)}{4+3A}$ & $A>-\frac{4}{3}$ & $-1-\frac{3A}{2}$ & $\frac{3(1+A)(1+3Ak\lambda)}{k}$ \\
$P_{12}$ & $\frac{2}{4+3A}$ & $\frac{2+3A}{4+3A}$ & $\frac{4(1+A)}{4+3A}$ & $A>-\frac{4}{3}$ & $-1-\frac{3A}{2}$ & $\frac{3(1+A)(1+3Ak\lambda)}{k}$ \\
  \noalign{\smallskip}\hline
 \end{tabular}
 \caption{Fixed points with existence condition and $q,\Omega_m$ for Kantowski-Sachs cosmologies}
 \label{table1}}      % Give a unique label
 \end{center}
 \end{table}
In summary, we may conclude following details about the fixed points:\\
Points $P_1$, $P_2$ exist for all values of variables and have eigenvalues $[2,-1,4+3A]$ and $[-2,1,-4-3A]$ respectively. These points 
are saddle in nature representing decelerating expansion with radiation dominated universe having effective equation of state 
$\gamma=\frac{1}{3}$. With $y=0$, these points lead to isotropization and thus represent isotropic universe. The matter density parameter 
depends on the value of Rastall parameter and will be positive only for $k\lambda<\frac{1}{4}$. These points are characterized by $x^2=1$, $y^2=0$, 
so the directional scale factors $a_1,a_2$ will evolve with same rate and $\Theta\propto \frac{3}{2t}$. \\
Points $P_3$, $P_4$, $P_5$ and $P_6$ have eigenvalues $[6,2,3(2+A)]$, $[-6,-2,-3(2+A)]$, $[2,2,3(2+A)]$ and $[-2,-2,-3(2+A)]$ respectively. $P_3$ 
and $P_5$ act as source (repeller) for $A>-2$ and saddle otherwise. $P_4$ and $P_6$ act as sink (attractor) for $A>-2$ and saddle otherwise. However, 
these points represent decelerating expansion of universe having domination of stiff matter like fluid $(\gamma=1)$. The matter density parameter 
at these points is zero. However, these points do not correspond to isotropic universe. With $\Omega_m=0$, these points correspond to the vacuum 
boundary solution. These points are characterized by $x^2=1$, $y^2=1$, and $\Theta\propto \frac{1}{t}$. Depending on the sign of $y$, the 
directional scale factors $a_1,a_2$ will take form either ($a_1\propto t$ and $a_2\propto$ constant) or ($a_1\propto t^{-\frac{1}{3}}$ and 
$a_2\propto t^{\frac{2}{3}}$).\\
Points $P_7$, $P_8$ have eigenvalues $[-3,-2,6(1+A)]$ and $[3,2,-6(1+A)]$ respectively. $P_7$ and $P_8$ act as sink and source respectively 
for $A<-1$ and saddle otherwise. These points represent decelerating expansion with domination of dust matter $(\gamma=0)$. The matter density 
parameter at these points is zero but these points correspond to the anisotropic universe. These points are characterized by $y^2=1$ and 
$\Theta\propto \frac{2}{3t}$. It is worth stressing that these two points are always placed outside the physical region of phase space.\\
Points $P_9$, $P_{10}$ have eigenvalues $\left[ -\frac{3}{2}(2+A),-2-3A,-4-3A\right] $ and $\left[ \frac{3}{2}(2+A),2+3A,4+3A\right] $ 
respectively. For $A>-\frac{2}{3}$, $P_9$ acts as attractor while $P_{10}$ is repeller and for $A<-2$, $P_9$ acts as repller 
while $P_{10}$ is attractor. Both points are of saddle type for  $-2<A<-\frac{2}{3}$. For $A=0$, the points exhibit de Sitter accelerated 
expansion. For super-accelerated expansion ($q<-1$), we need $A>0$. With $A<0$, we get $q>-1$. In particular, $q<0$ for 
$A\in \left(-\frac{2}{3},0 \right)$, $q=0$ at $A=-\frac{2}{3}$ and $q>0$ for $A<-\frac{2}{3}$. These points represent $\gamma>0$ ($<0$) 
for $A<-1$ ($>-1$) respectively. The matter density parameter at these points depends on Rastall parameter and $\alpha$. These points 
correspond to isotropic universe. These points are characterized by $x^2=1$ and for $A=-\frac{4}{3}$, which is less than $-1$, these points 
will coincide with $P_1$ and $P_2$ respectively. For these points, $\Theta\propto -\frac{2}{At}$. For $q=-1$ scenario which is possible at 
$A=0$, $\Theta\propto $ constant and $a_1,a_2 \propto e^{\pm \Theta t}$ depending on the sign of $x$. Condition for $A<-2$ with 
$0\leq \alpha\leq 1$ may be written in terms of $k\lambda$ as $\left(0\leq \alpha <\frac{1}{3}\land \frac{1}{3 \alpha +3}<k\lambda <\frac{1-\alpha }
{2 \alpha +2}\right)\lor \left(\frac{1}{3}<\alpha \leq 1\land \frac{1-\alpha }{2 \alpha +2}<k\lambda<\frac{1}{3 \alpha +3}\right)$.\\
Points $P_{11}$ and $P_{12}$ exist for $A>-\frac{4}{3}$ and the eigenvalues of these points have complicated and long expression of $A$, 
so we omit to write here. Stability nature of these points may be classified on the basis of $k\lambda$ and $\alpha$. These points will represent 
accelerated expansion for $A\in \left(-\frac{2}{3},0 \right) $, $q=0$ at $A=-\frac{2}{3}$ and decelerated expansion for $A<-\frac{2}{3}$. These 
points may represent $\gamma>0$ ($<0$) for $A<-1$ ($>-1$) respectively. The matter density parameter at these points depends on Rastall 
parameter and $\alpha$ but, these solutions correspond to the anisotropic universe. These points may be characterized by $x^2=1$ and $u=0$ for 
$A=-\frac{2}{3}$ and $A=-1$ respectively. For these points, $\Theta\propto -\frac{2}{At}$. For $q=-1$, $\Theta\propto $ constant and these 
points will realize de Sitter expansion.\\
The system (\ref{eq6}) has three invariant sub-manifolds characterized by $x=1,x=-1$ and $u=0$. On $u=0$ sub-manifold, the system is independent 
of $A$. On $u=0$, the system takes the form
 \begin{eqnarray}
 x'=(x^2-1)\left(1+xy+y^2\right), \quad
 y'=(1-x^2)(1-y^2)+xy\left(y^2-1 \right)
 \label{eq6b}
 \end{eqnarray}
Points $P_i,i=1,2,3,4,5,6$ belong to subspace $u=0$. The phase space in $x-y$ plane has been given in Fig. (\ref{figure1}). 
Points $P_1,P_3,P_5,P_9$ belong to the expanding cosmological space $x=1$ and  points $P_2,P_4,P_6,P_{10}$ belong to the contracting cosmological space $x=-1$. The behavior of $(u,y)$ space have been given in Fig. (\ref{figure2}) for $x=1$ and 
Fig. (\ref{figure3}) for $x=-1$. We have taken $A=-2.2,-1.55,-1,-0.6$ as representative values of $A$ in region $A<-2$, $-2<A<-\frac{4}{3}$, $-\frac{4}{3}<A<-\frac{2}{3}$, $A>-\frac{2}{3}$ respectively. Note that 
$k\lambda=\frac{1+\alpha+A}{(4+3A)(1+\alpha)}$ and $0\leq \alpha\leq 1$ are constraints for classical stability of the model. On $x=\pm 1$, we have
  \begin{eqnarray}
  y'=\pm y\left(y^2-1-\frac{3}{2}u \right), \quad   u'=\pm u\left(4+3A-3u+2y^2\right)
  \label{eq6a}
  \end{eqnarray}
\section{LRS Bianchi I and Bianchi III cosmology}
\label{LRSB}
For LRS Bianchi I and Bianchi III model, we perform dynamical system analysis in an unified way. In order to transform Eq. (\ref{eq2}) 
into autonomous system, we define the dynamical variables 
\begin{equation}
x=\frac{k(\rho+p)}{\Theta^2}, \ \ y=\frac{\sigma}{\Theta}, \ \ z=\frac{{}^3R}{2\Theta^2}, \ \  v=\frac{k\rho}{\Theta^2}
\label{eq8}
\end{equation} 
In this section, `prime' denotes derivative with respect to $\tau$ and $d\tau=\frac{\Theta}{3}dt$. 
Also, $-1\leq x\leq 1$, $-1\leq y\leq 1$, $-1\leq z\leq 0$ and $v\geq 0$. For the LRS BI model, ${}^3R=0$ and thus in terms of dynamical 
variables, the invariant set $z=0$ corresponds to the LRS BI case. One may also have some critical points with $z=0$ in LRS BIII cosmology 
that belongs to the BI boundary. In this situation, stability in case of Bianchi III model will be different because, the system can evolve 
in $z$-direction, which is not the case in Bianchi I model \cite{crf}. In terms of dynamical variables, we have $3v=9k\lambda x+(1-4k\lambda)
(1+3z-3y^2)$ and the deceleration parameter $(q)$, matter density parameter ($\Omega_m$) may be written as  
\begin{eqnarray}
q=1+3z+3y^2-\frac{9}{2}x, \quad
\Omega_m=\frac{\rho}{3H^2}=\frac{9k\lambda x+(1-4k\lambda)(1+3z-3y^2)}{k}
\label{eq9}
\end{eqnarray}
Using the auxiliary relation between $x,y,z$ and $v$, we may reduce the dimension of phase space and thus, the $3$-dimensional autonomous 
system may be written as
\begin{eqnarray}
x'=x\left(4+3A-9x+6y^2+6z \right) \nonumber \\
y'=3y^3+\sqrt{3}z+y\left( -1+3z-\frac{9}{2}x\right) \nonumber \\
z'=z\left(2-9x+2\sqrt{3}y+6y^2+6z \right) 
\label{eq10}
\end{eqnarray}
We use $p=f(\rho)$ and $f'(\rho)=\frac{df}{d\rho}$. For $p=\alpha\rho$, we find the fixed points of system by setting $x'=0,y'=0,z'=0$. 
For the above system (\ref{eq10}), there are $7$ fixed points. Various details about these fixed points have been listed in Table (\ref{table2}). 
The fixed points which correspond to expansion, moreover are stable representing isotropic universe may represent the late-time state of the universe. 
Using definition of the deceleration parameter, we may have $a(t)\propto t^{\frac{1}{1+q_0}}$, where $a(t)$ is mean scale factor related to 
volume scale factor $V$ given by $V=a^3=a_1{a_2}^2$ and $q_0$ is the deceleration parameter value calculated at the fixed point.  \\ 
 \begin{table}[h!]
 \begin{center}
 {\begin{tabular}{cccccc}
 \hline\noalign{\smallskip}
 Point & $x$ & $y$ & $z$ &  $q$ & $\Omega_m$ \\
 \noalign{\smallskip}\hline\noalign{\smallskip}
$Q_1$ & $0$ & $0$ & $0$ & $1$ & $\frac{1-4k\lambda}{k}$ \\
$Q_2$ & $0$ & $\frac{1}{\sqrt{3}}$ & $0$ & $2$ & $0$ \\
$Q_3$ & $0$ & $-\frac{1}{\sqrt{3}}$ & $0$ & $2$ & $0$ \\
$Q_4$ & $0$ & $-\frac{1}{\sqrt{3}}$ & $-\frac{1}{3}$ & $1$ & $\frac{4k\lambda-1}{k}$ \\
$Q_5$ & $0$ & $-\frac{1}{2\sqrt{3}}$ & $-\frac{1}{4}$ & $\frac{1}{2}$ & $0$ \\
$Q_6$ & $\frac{1}{9}(4+3A)$ & $0$ & $0$ & $-1-\frac{3A}{2}$ & $\frac{(1+3Ak\lambda)}{k}$ \\
$Q_7$ & $\frac{1}{3}(4+7A)+A^2$ & $\frac{1}{2\sqrt{3}}(2+3A)$ & $1+2A+\frac{3}{4}A^2$ & $-1-\frac{3A}{2}$ & $\frac{3(1+A)(1+3Ak\lambda)}{k}$ \\
\noalign{\smallskip}\hline
 \end{tabular}
 \caption{Fixed points with existence condition and $q,\Omega_m$ for LRS-BIII and BI cosmologies}
 \label{table2}}      % Give a unique label
 \end{center}
 \end{table}
\begin{figure}[h!]
	\centering
	\includegraphics[width=.425\textwidth]{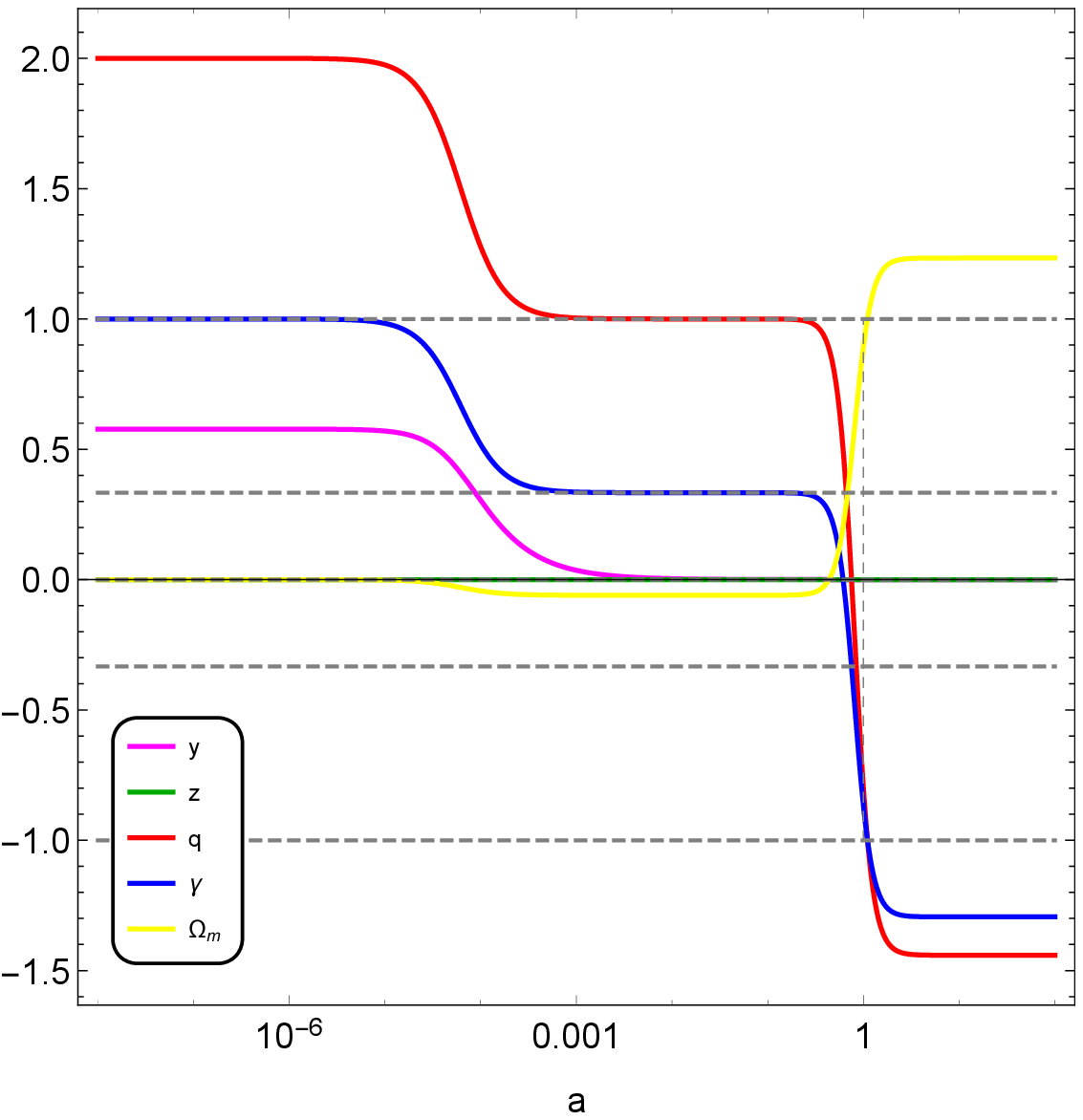}\hfill
	\caption{Cosmological parameters with mean scale factor $a$}
	\label{figure4}
\end{figure}
 In summary, we may conclude following details about the fixed points:\\
Point $Q_1$ exists always and has eigenvalues $[2,-1,4+3A]$, which means that the point is saddle in nature. The point represents 
 isotropic, decelerating universe dominated by radiation $(\gamma=\frac{1}{3})$ component. This point exists for both LRS BIII and BI geometries. 
 This point is characterized by $y^2=0$ and $z=0$, so the directional scale factors $a_1,a_2$ will evolve with same rate and 
 $\Theta\propto \frac{3}{2t}$.\\ 
 Points $Q_2$ and $Q_3$ have eigenvalues $[6,2,3(2+A)]$ and $[2,2,3(2+A)]$ respectively. These points are saddle in nature for $A<-2$ and 
 unstable otherwise; represent anisotropic, decelerating universe dominated by stiff matter fluid $(\gamma=1)$. These points exist for both 
  LRS BIII and BI geometries. These points are characterized by $x=0$, $z=0$, and $\Theta\propto \frac{1}{t}$. Depending on the sign of $y$, the 
  directional scale factors $a_1,a_2$ will take form either ($a_1\propto t$ and $a_2\propto$ constant) or ($a_1\propto t^{-\frac{1}{3}}$ and 
  $a_2\propto t^{\frac{2}{3}}$).\\
Point $Q_4$ has eigenvalues $[-2,1,4+3A]$. The point has a saddle nature and represents an anisotropic, decelerating universe dominated by radiation. 
The point has $z<0$ and thus represents LRS BIII geometry. This point is characterized by $y^2\neq 0$ and $z<0$, so the directional scale 
factors $a_1,a_2$ do not evolve with the same rate; however, the volume expansion will be characterized by effective radiation-like 
fluid with $\Theta\propto \frac{3}{2t}$.\\
Point $Q_5$ has eigenvalues $[-\frac{3}{2},-1,3(1+A)]$. The point acts as attractor for $A<-1$ and saddle otherwise. The point represents 
anisotropic, decelerating universe dominated by dust matter having LRS BIII geometry. This point follow $\Theta\propto \frac{2}{3t}$.\\ 
Point $Q_6$ has eigenvalues $[-\frac{3}{2}(2+A),-2-3A,-4-3A]$. 
The point is attractor for $A>-\frac{2}{3}$, repeller for $A<-2$ and saddle for $-2<A<-\frac{2}{3}$. The point represents isotropic universe 
dominated by fluid having $\gamma>0$ ($<0$) for $A<-1$ ($>-1$) respectively. The point will represent accelerated universe for 
$A\in \left(-\frac{2}{3},0 \right) $, $q=0$ at $A=-\frac{2}{3}$ and decelerated universe for $A<-\frac{2}{3}$. The matter density parameter 
at this point depend on Rastall parameter. This point exists for both LRS BIII and BI geometries. This point is characterized by $y^2=0$ and 
$z=0$ and for $A=-\frac{4}{3}$, which is less than $-1$, this point will coincide with origin of phase space. We have $\Theta\propto -\frac{2}{At}$. 
For de Sitter expansion scenario which is possible at $A=0$, $\Theta\propto $ constant and $a_1,a_2 \propto e^{\pm \theta t}$.\\
Point $Q_7$ has eigenvalues $\left[ -4-3A,\frac{3}{4}\left(-2-A\pm\sqrt{36+100A+89A^2+24A^3}\right)\right] $.  Stability nature of 
this point may be classified on the basis of $k\lambda$ and $\alpha$. The point will represent accelerated universe for 
$A\in \left(-\frac{2}{3},0 \right) $, $q=0$ at $A=-\frac{2}{3}$ and decelerated universe for $A<-\frac{2}{3}$. The point will represent 
anisotropic universe dominated by fluid having $\gamma>0$ ($<0$) for $A<-1$ ($>-1$) respectively. The point exists in general for LRS BIII 
geometry but may also exist for LRS BI geometry for $A=-2,-\frac{2}{3}$. This point may be characterized by $x^2=1$ and $u=0$ for 
$A=-\frac{2}{3}$ and $A=-1$ respectively. For this point $\Theta\propto -\frac{2}{At}$. For different values of $A$, this point may belong 
to $x=0$ or $z=0$ family. \\
The behavior of cosmological parameters like deceleration parameter and equation of state parameter $\gamma$ along with other relevant cosmological 
quantities have been displayed in Fig. (\ref{figure4}) for $\alpha=0.001,k=1,\lambda=0.265,A=0.294116$. The parameter's values adopted here are 
primarily motivated to get desirable late-time accelerating, isotropic universe evolution. The value of $\lambda$ is far away from its constrained 
value in Rastall gravity with flat FRW spacetime \cite{51a,23y} but are consistent with Batista et al. \cite{23z}. The qualitative analysis of the LRS BI and LRS BIII model shows that there are 
two fixed points $Q_1$ and $Q_4$ which may yield negative matter density (alternatively) in expanding cosmology on the basis of 
$k\lambda>\frac{1}{4}$ or $k\lambda<\frac{1}{4}$. The negative matter density may also be visualized as the violation of weak energy conditions. 
The model may display a late-time accelerating universe with isotropic evolution at present times, which is compatible with observations 
\cite{planck2013,Pl2018}, but may violate the weak energy condition in the past. The model exhibit solutions having stiff equation of state like evolution during initial times due to the presence of anisotropy. Dissipation of anisotropy may leads to the evolution of universe into radiation dominated era followed by matter dominated era (see Fig. (\ref{figure4})). It is worthwhile to mention that the model may evolve into 
phantom territory $(\gamma<-1)$ in the future with $\Omega_m>1$ evolving towards a finite-time future singularity \cite{m2019}.\\
\begin{figure}[h!]
	\centering
	\includegraphics[width=.4\textwidth]{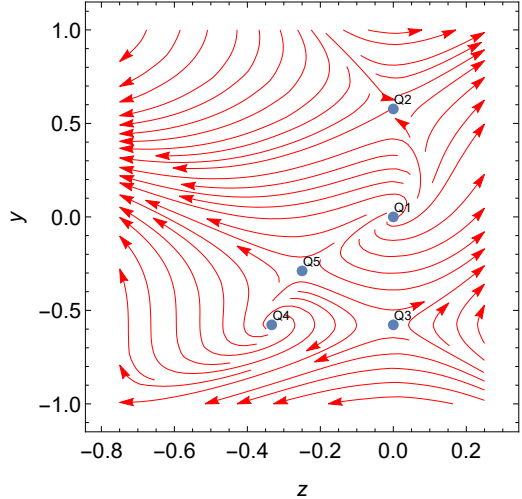}\hfill
	\caption{Phase space $(z,y)$ in the sub-manifold $x=0$}
	\label{figure5}
\end{figure}
\begin{figure}[h!]
	\centering
	\subfigure[]{\includegraphics[width=0.375\textwidth]{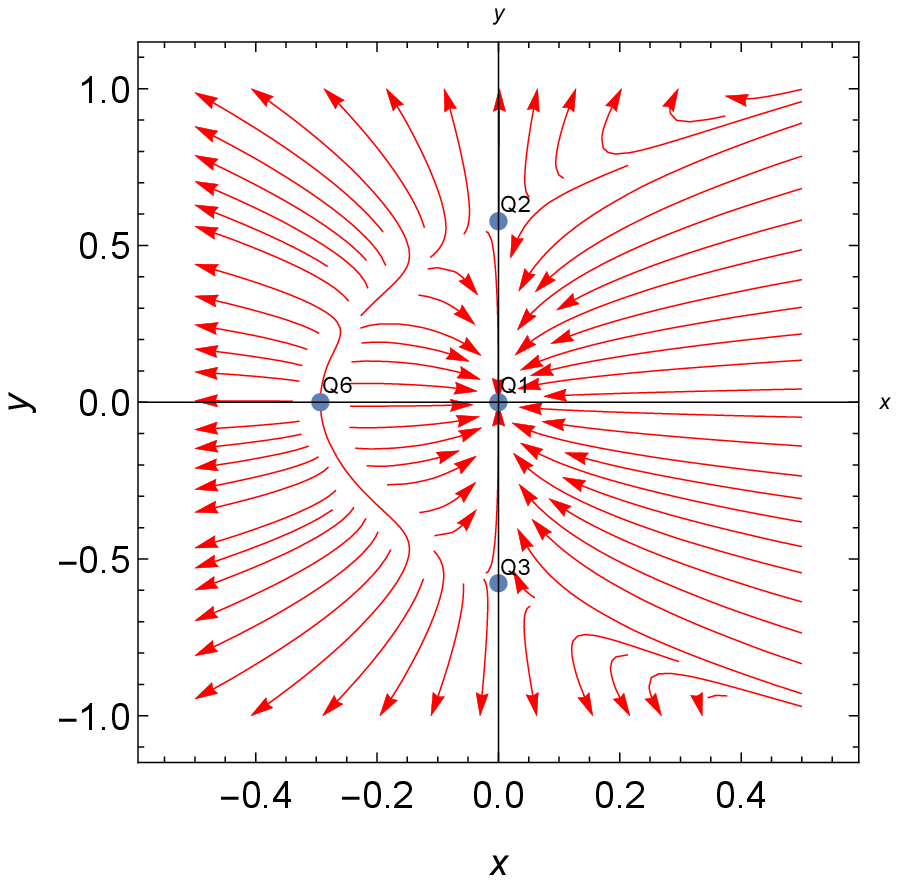}} \hfill
	\subfigure[]{\includegraphics[width=0.375\textwidth]{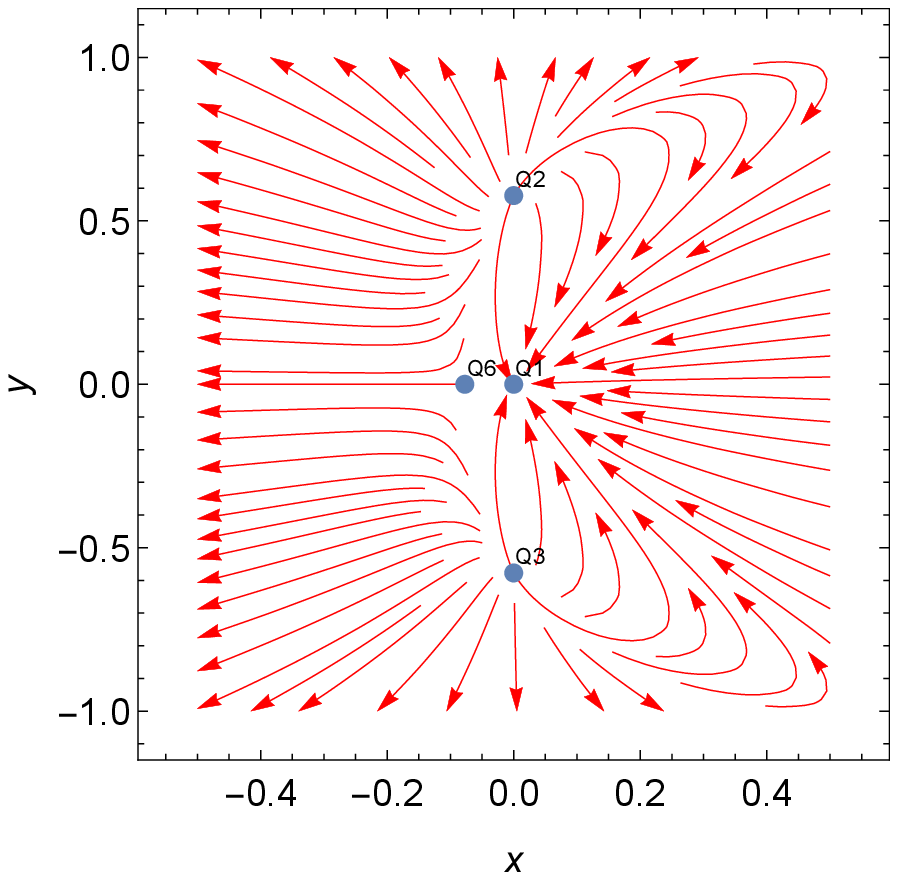}} 
	\subfigure[]{\includegraphics[width=0.375\textwidth]{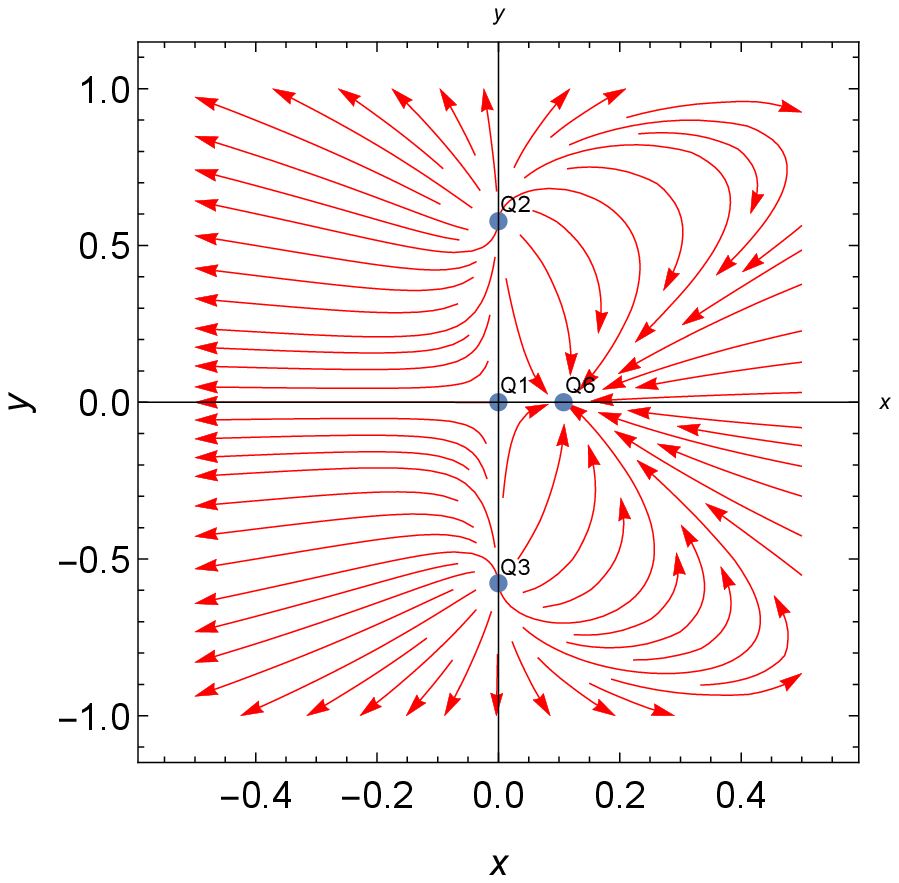}}\hfill
	\subfigure[]{\includegraphics[width=0.375\textwidth]{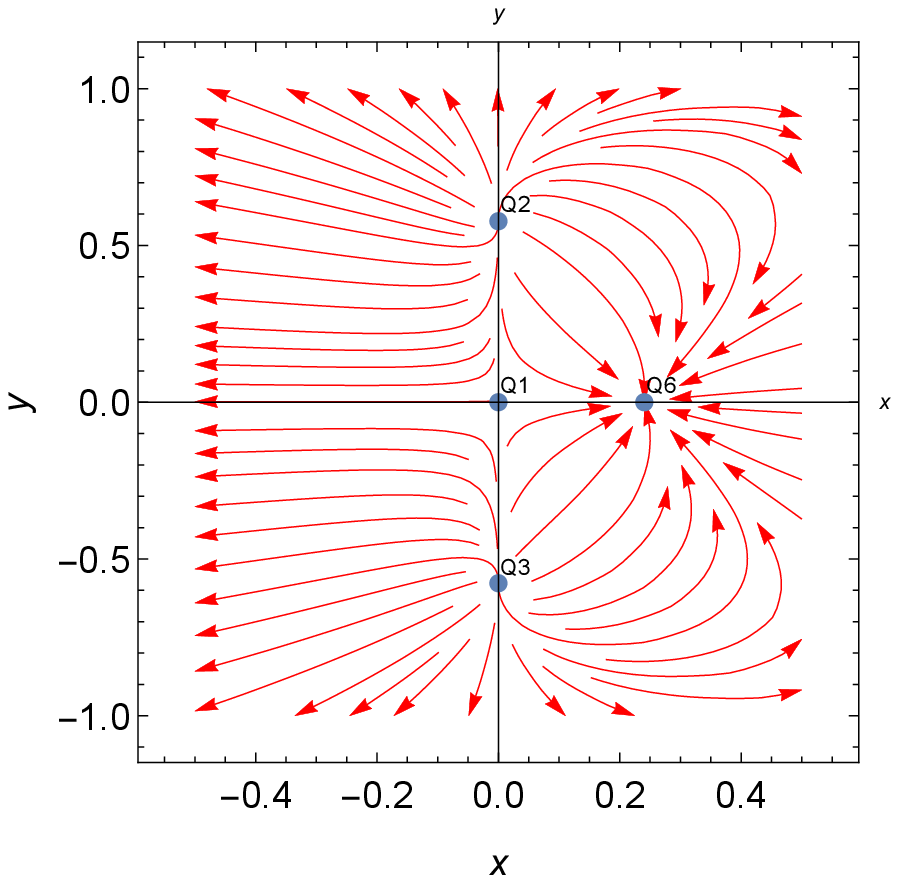}}
	\caption{Phase space $(x,y)$ in the sub-manifold $z=0$ for (a) $A<-2$, (b) $-2<A<-\frac{4}{3}$, (c) $-\frac{4}{3}<A<-\frac{2}{3}$ 
	and (d) $A>-\frac{2}{3}$ respectively \label{figure6}}
\end{figure}
The system (\ref{eq10}) has two invariant sub-manifolds characterized by $x=0$ and $z=0$. On $x=0$, we have
 \begin{eqnarray}
y'=3y^3+\sqrt{3}z+y\left(3z-1\right), \quad  z'=z\left(2+2\sqrt{3}y+6y^2+6z \right) 
 \label{eq10a}
 \end{eqnarray}
Points $Q_i,i=1,2,3,4,5$ belong to $x=0$ subspace. This subspace is independent of values of $A$. The phase space behviour in $(z,y)$ 
plane is given in Fig. (\ref{figure5}). On $z=0$ subspace, we have
 \begin{eqnarray}
 x'=x\left(4+3A-9x+6y^2\right),  \quad  y'=3y^3-y\left(1+\frac{9}{2}x\right) 
 \label{eq10b}
 \end{eqnarray}
 This subspace corresponds to the LRS Bianchi-I cosmologies where ${}^3R=0$. The behaviour in $(x,y)$ phase space is given in Fig. (\ref{figure6}). 
 We have taken $A=-2.2,-1.55,-1,-0.6$ as representative values of $A$ in region 
 $A<-2$, $-2<A<-\frac{4}{3}$, $-\frac{4}{3}<A<-\frac{2}{3}$, $A>-\frac{2}{3}$ respectively.

\section{General issues}
\label{GI}
\subsection{Bounce conditions}
\label{bc}
In cosmological modeling, bouncing models are of special interest. Bouncing universe undergoes a collapse, attains
a minimum, and then subsequently expands during their evolution \cite{sm2020}. In anisotropic space-time, we may define the occurrence of 
bounce at time $t=t_b$ by \cite{Tra2015a,Tra2015b,Tra2017}
\begin{enumerate}
\item $\Theta(t_{b})=0 $ 
\item $ \dot{\Theta}(t)>0$ for $t\in[t_{b}-\epsilon, t_{b})\cup (t_{b},t_{b}+\epsilon]$ for small $\epsilon >0$
\end{enumerate}  
However, anisotropic models have more than one scale factors $a_{i}$, $i=1,2$, the above conditions of bounce should be understood as 
characterizing a bounce in average scale factor $V=a_{1}{a_2}^2$. One may also consider a more generic situation where a bounce can 
occur in any of the directional scale factors $a_{i}$. We can make this more precise by defining the expansion parameters 
$H_{i}=\frac{\dot a_{i}}{a_{i}}$. So a bounce in $a_{i}$ will occur at $t=t_{b}$  if
\begin{enumerate}
\item $H_{i}(t_{b})=0$ 
\item $\dot{H_{i}}(t_{b})>0$ for $t\in[t_{b}-\epsilon, t_{b})\cup (t_{b},t_{b}+\epsilon]$ for small $\epsilon >0$
\end{enumerate}
Although it may be possible to have a bounce in any one of the $a_i$s but not the other, this does not lead to a new expanding universe 
region \cite{dspks2006}. In this paper, we consider the bounce in average scale factor.\\
From Eq. (\ref{eq2}), we may write 
\begin{eqnarray}
k\rho=2k\lambda \dot{\Theta}+\frac{1}{3}\Theta^2+\frac{1}{2}(1-2k\lambda)\cdot {}^3R+(6k\lambda-1)\sigma^2\nonumber \\
kp=\frac{2}{3}(1-3k\lambda) \dot{\Theta}+\frac{1}{9}\Theta^2+\frac{1}{2}(6k\lambda-1)\cdot {}^3R+(5-18k\lambda)\sigma^2
\label{eq11}
\end{eqnarray}
Above expressions may yield various linear combinations of $\rho$ and $p$ which are known as energy conditions. These may categorize certain 
physically reasonable ideas in a precise manner. These point-wise conditions are coordinate invariant restrictions on stress-energy tensor 
of model $T_{ij}=\text{diag}(\rho,-p,-p,-p)$ and may be given by \cite{sm2020}
\begin{enumerate}
\item Null energy condition (NEC) $ \Leftrightarrow \rho + p\geq 0 $
\item Weak energy condition (WEC) $ \Leftrightarrow \rho \geq 0,\ \rho + p\geq 0  $
\item Dominant energy condition (DEC) $\Leftrightarrow \rho\geq 0,\ \rho - p\geq 0, \rho +p\geq 0 $
\item Strong energy condition (SEC) $\Leftrightarrow \rho + p\geq 0,\ \rho + 3p\geq 0 $
\end{enumerate}
These conditions may also be compatible with cosmic acceleration provided that a component yielding repulsive gravity exists in model and 
acceleration stays within certain bounds. In study of cosmological bounce, these conditions are of special interest. From Eq. (\ref{eq11}), 
at the bounce instant $t=t_b$, we may have
\begin{eqnarray}
k\rho=2k\lambda \dot{\Theta}+(6k\lambda-1)\sigma^2+\frac{1}{2}(1-2k\lambda)\cdot {}^3R\nonumber \\
k(\rho+p)=\frac{2}{3}\dot{\Theta}+\frac{2}{3}\sigma^2+\frac{1}{3}\cdot {}^3R\nonumber \\
k(\rho+3p)=2(1-2k\lambda)\dot{\Theta}+4(1-3k\lambda)\sigma^2+2k\lambda \cdot {}^3R
\label{eq12}
\end{eqnarray}
From Eq. (\ref{eq12}), we may have following inferences at and near the bounce point $t=t_b$:
\begin{enumerate}
\item In KS model, ${}^3R>0$, therefore
\begin{enumerate}
\item $k\rho>0 \Rightarrow \frac{1}{6}<k\lambda<\frac{1}{2}$,
\item $k(\rho+p)>0  \ \ \forall \ \ k\lambda$,
\item $k(\rho+3p)>0 \Rightarrow 0<k\lambda<\frac{1}{3}$.
\end{enumerate}
\item In LRS BI model, ${}^3R=0$, therefore
\begin{enumerate}
\item $k\rho>0 \Rightarrow k\lambda>\frac{1}{6}$,
\item $k(\rho+p)>0  \ \ \forall \ \ k\lambda$,
\item $k(\rho+3p)>0 \Rightarrow k\lambda<\frac{1}{3}$.
\end{enumerate}
\item In LRS BIII model, ${}^3R<0$, therefore
\begin{enumerate}
\item $k\rho>0 \Rightarrow k\lambda>\frac{1}{2}$,
\item $k(\rho+p)>0$ whenever $\dot{\Theta}+\sigma^2$ will be dominating upon $\frac{{}^3R}{2}$,
\item $k(\rho+3p)>0 \Rightarrow k\lambda<0$.
\end{enumerate}
\end{enumerate}
In the above discussion, we take $k>0$. For $k<0$, the above conditions will be reversed. These conditions provide significant deviations 
from LRS models of general relativity where a cosmological bounce is not permitted unless the reality condition for momentum density is violated. 
The non-minimal coupling parameter $\lambda$ provides opportunity for different models of Rastall gravity to have bounce and/or cyclic scenarios. 
Different bouncing and cyclic solutions in Rastall gravity have been exhibited in the literature using analytical as well as dynamical system methods 
\cite{9,sm2020,01,1a}.
\subsection{Isotropization} 
\label{lti}
In an expanding universe, the late-time isotropization can be characterized by vanishing shear, or alternatively we can use the condition 
$\frac{\sigma}{\Theta}\rightarrow 0$ as $t\rightarrow \infty$ \cite{coll,crf}. For expanding LRS BI and BIII geometries, these criterion 
may be given by $H\geq 0$, $\dot{H}\leq 0$ and $\sigma, \rho+p,{}^3R\rightarrow 0$ with $H\rightarrow$ constant (small) as 
$t\rightarrow \infty$ \cite{sbds}.\\
We use the dynamical system analysis to extract the asymptotic behavior of LRS models by explicitly identifying the isotropic/anisotropic 
late-time stable fixed points.\\
For the Kantowski-Sachs model, point $P_9$ represents a stable, isotropic, expanding solution which is a late-time attractor. On the sub-manifold 
$x=\pm 1$, point $P_1$ represents an isotropic solution with the universe having effective radiation matter expanding with deceleration. Point 
$P_{10}$ may also represent a stable contracting solution that is isotropic; however, depending on the value $A$, this point may represent 
an unstable expanding solution that is isotropic in nature.\\
In LRS Bianchi models, point $Q_5$ may act as anisotropic stable solution having a matter-dominated expansion. Point $Q_6$ is a desirable 
isotropic attractor solution that may represent asymptotically de Sitter expansion at late times.\\
The expanding, accelerating, late-time attractor solutions are possible in Kantowski-Sachs, LRS Bianchi I, and Bianchi III geometries which 
may be isotropic on the basis of parameter value $A$ in the finite region of the phase space.  
\section{Conclusions}
\label{conclusion}
We have considered a detailed dynamical system analysis of the Rastall cosmological model with anisotropic geometries and, in particular, Kantowski-Sachs, 
LRS Bianchi I and Bianchi III cases. We have shown that the governing cosmological equations may reduce into a simple three-dimensional autonomous 
dynamical system. These geometries may provide the late-time isotropized solutions with an explanation for observed universe isotropy instead of
having to assume it from the beginning, as in FRW cosmologies. These isotropized solutions exhibit observables (like deceleration parameter and 
effective EoS) in agreement with observations, independently of the initial conditions and their specific evolution. The universe model can also 
result in dark-energy dominated, accelerating, de Sitter solution at late times on the basis of the value of model parameters.\\
We have shown that in examined geometries with Rastall gravity, it is possible to have a cosmological bounce satisfying the null energy condition. 
However, the constraints on model parameters may be different in the considered geometries. This result is in contrast with the corresponding 
anisotropic model of General Relativity, where the cosmological bounce is possible only by violation of the null energy condition.\\
 We have investigated the directional scale factors at the fixed points. The late-time behavior of the model may have anisotropic 
collapsing solution of Kasner-like type or exponentially expanding solutions of de Sitter-like type. It would be interesting to investigate 
the analytical solutions of field equations in the considered model. We leave it for a future study.
\section*{Acknowledgments}
We are thankful to the anonymous reviewer for the illuminating remarks, which have been helpful to improve the presentation of the manuscript. 
A. Pradhan also thanks the IUCAA, Pune, India for providing facility and support under visiting associateship program.

\end{document}